# In Search of Outstanding Research Advances
## Prototyping the creation of an open dataset of "editorial highlights"

Alexis-Michel Mugabushaka[1], Jasmin Sadat[1] and Jorge Costa Dantas Faria[2]

[1]European Research Council Executive Agency (ERCEA)[1], Covent Garden Building, Place Charles Rogier 16, 1210 Bruxelles (Belgium)

[2] Joint Research Centre (JRC)[2]

**Abstract**

A long-standing research question in bibliometrics is how one identifies publications, which represent major advances in their fields, making high impact in there and other areas. In this context, the term "Breakthrough" is often used and commonly used approaches rely on citation links between publications implicitly positing that peers who use or build upon previously published results collectively inform about their standing in terms of advancing the research frontiers.

Here we argue that the "Breakthrough" concept is rooted in the Kuhnian model of scientific revolution which has been both conceptually and empirically challenged. A more fruitful approach is to consider various ways in which authoritative actors in scholarly communication system signal the importance of research results. We bring to discussions different "recognition channels" and pilot the creation of an open dataset[3] of editorial highlights from regular lists of notable research advances. The dataset covers the last ten years and includes (a) the "discoveries of the year" namely: Science - Breakthroughs of the year and La Recherche - les 10 découvertes de l'année and (b) weekly editorial highlights from Nature ("research highlights") and Science ("editor's choice"). The final dataset includes 230 entries in the "discoveries of the years" (with over 720 references) and about 9,000 weekly highlights (with over 8,000 references).

**Introduction**

By most popular accounts in the history of science, scientific advances seem to follow an apparently easily described course. Discoveries (and inventions) built upon those from previous work or previous generations, pushing the frontiers of the unknown (and the possible) and along the way fundamentally changing our perception of the world around us (and in us) and leading to applications which changes our lives (mostly to the better).

Historians of science and scholars of its evolutionary patterns have however often warned against this simplistic model of growth of science.

Thomas Kuhn (1962) was arguably the first to challenge this view of scientific advances as a linear, cumulative, and progressive process towards the end-truth. In his "*structure of scientific revolutions*", he argued that scientific advances happen in "jumps" which, oftentimes, require the revision (and repudiation) of existing scientific tenets and practices. Between those jumps lie periods of "*normal science*" in which researchers work on more or less pre-defined research questions using established concepts and methods. Thomas Kuhn argued that as incremental results accumulate, they unavoidably uncover facts which may not be reconcilable with the established understanding, leading to small and major crises in existing theories. This persists until new models arise which can account for those "anomalies". Kuhn coined the term "*paradigm shift*" to describe this transition and "revolutionary science" as "*the relatively brief period after a paradigm crisis (which) comes to an end once a new paradigm is installed.*" (Fuller, p. 224).

---

"*Paradigm shift*" went on to become one of the most important concepts in the studies of scientific advances (and beyond). The concept has not only decisively shaped the view of how science works and grows, but also significantly impacted (and decisively guided) science policy debates of the last decades, especially debates about which type of research to prioritise in funding . For example, it is the basis of the term "*transformative research*" that the NSF popularised in research funding circles in the mid-2000s as *"(..) discussions of Transformative research typically assume a Kuhnian framework."* (Frodeman et a. 2012, p. 4). Transformative is defined as "*research driven by ideas that have the potential to radically change our understanding of an important existing scientific or engineering concept or leading to the creation of a new paradigm or field of science or engineering. Such research also is characterized by its challenge to current understanding or its pathway to new frontiers* (NSB 2007, p. 10).

A concept related to "transformative research" is "*Breakthrough*". In its most common definition - as "*a discovery (an observation or finding of something previously unknown) showing a major impact on future scientific research and considered to signal possible breaches, focus shifts, or even turning points in science*" (Winnink 2017, p.1) – it clearly shows its deep roots in Kuhn's model of scientific advance. This concept also plays an important role in science policy debates as shown by *"(the) attention (…) given at the national and international levels to setting up funding instruments and research infrastructures with the intention to motivate and foster breakthrough research*" (Schneider et al. 2017, p. 2).

Despite its success - also beyond the studies of science - the Kuhn model of science advancing through paradigm shifts was controversial as soon as it was published. One of the earliest critics was Karl Popper, who argued that the distinction between "normal science" and "revolutionary science" is not as "quite sharp as Kuhn makes it" (Popper, 1970, p. 52). In his view, "*few, if any, scientists who are recorded by the history of science were "normal*" scientists in Kuhn's sense" (p. 52). While in Kuhn's view "*many great scientists must have been "normal"*, Popper's view is that "*science is in permanent revolution*".

Toulmin (1970) took this argument further and developed an alternative to the Kuhnian model of scientific revolution. Asking if "*the distinction between normal and revolutionary science holds water*", he seizes on the metaphor of revolution and points out that scholarship of political revolutions has shown that, other than what the term implies, revolutions in fact never means breach of continuity. Historic examples of political revolutions show that" continuities *have been as important as changes*", the differences between revolution changes and normal changes being only a matter of degree. Transposing this idea to scientific progress, he criticized the model of Kuhn of going too far in its focus of discontinuities in scientific theories (between paradigms). He argues that as no conceptual change in science is ever absolute, the central idea of "incommensurability" in Kuhn's model (the idea that paradigms have little in common as they exist in different spheres) is not tenable. However an even more serious charge against the "revolutionary model" is that, it dismisses important discoveries and developments which happen within a "paradigm" (i.e not as result of anomalies) but which themselves can spark radical changes in their fields and beyond. Toulmin sees those "micro-revolutions" as far too common than paradigm shifts would imply and therefore as the normal "units of all scientific innovation" to which those studying evolution of science should focus. On this line of argument, Casadevall and Fang (2016) remarked in their editorial on "*revolutionary science*", that by seeking to make a distinction between "normal and revolutionary science, the Kuhnian model "*does not apply to what is perhaps the most important biological finding in the 20th century*" - *the discovery of the structure of DNA -which "resulted from puzzle solving and normal science without any paradigmatic crisis* (p. 4)."

There have been attempts to empirically assess which model better accounts for some of the most important discoveries in recent times with mixed results.

Analyzing about 15 million references to papers published in 1998 and indexed in Web of Science, Van Raan (2000) finds evidence that science is a largely self-organizing system. He observed that in this system "*smaller and larger discoveries are the 'rule' ... it is impossible to regard the larger discoveries as essentially different, in terms of the system dynamics, from smaller breakthroughs.*" In his view this challenges Kuhn's model of scientific development. "*There is no 'normal' science alternated with well-defined periods of 'revolutionary science' in which new paradigms start (...) Science is always revolutionary*. (p.360).

Marx and Bornman (2016) used bibliometric methods to study the transition to the Big Bang Theory in cosmology and found the development not to be consistent with the Kuhn model. They concluded rather that there was no "*single scientific revolution in cosmology, but instead a paradigm shift that progressed as a slow, piecemeal process.*" In their view their results can be applied to other scientific developments. They observed that "*Within the history of modern science it is difficult to find examples that confirm Kuhn's model of comparatively long "normal" periods of unspectacular "dying up" work, interrupted only every once and a while by revolutions.*"

The editorial of Casadevall and Fang (2016) cited above surveyed a selection of what they see as revolutionary scientific advances and come to the conclusions that "*scientific revolutions have developed in a variety of ways. (...) some indeed seems to correspond to Kuhn's description (...) however many others do not correspond to (it).*"

This discussion on the models of scientific changes raises an important question: what are the research discoveries which significantly advance science?

If they are not those which trigger a paradigm shift (or possibly also those exposing the limitations of previous paradigms) as the Kuhn model would suggest, but include also those happening within "paradigms", how can they be identified? In other words, are there reliable ways to tell, from the multitudes of scientific advances made by researchers which ones stand out by the exceptional contribution they make to advance in science?

It is obvious that all the terms used to describe those advances are ambiguous. Importance, being a multifaceted and multidimensional characteristic, is a matter of perspective. As Casadevall and Fang (2009) remind us "*revolutionary science is not synonymous with extra-ordinary or important science*" (p. 2) and conceptual novelty is not necessarily an indication of importance. "*Discoveries may have enormous practical utility even though they are not conceptually new" (Casadevall and Fangl 2009, p. 4177)*.

From the obvious observation that research advances are "important" because some (authoritative) actors in the research communication system recognised them as such (from their perspective), important advances can be defined (imprecisely) as those standing out in the views of different actors. Identifying the different "recognition channels" and collecting the advances they list, could help the efforts to detect outstanding advances.

One such recognition channel are the "editorial highlights".

The objectives of this paper are:

- to reflect on the role of "editorial highlights" in the science recognition system and
- to discuss the feasibility of the creation of open, community-maintained data resources to record editorial highlights for use in bibliometric research and evaluative bibliometric.

## 2. Recognition as Landmark of outstanding advances in science

Simply put, science advances occur when previously unknown phenomena are discovered and added to its body. Implicit in this definition is the originality principle, one of the core norms of science (Merton 1973). It is "*through originality, in greater or smaller increments, that knowledge advances*" (...) and originality is one of the foundations of the science reward system in which "*recognition and esteem accrue to those who have made genuinely original contributions to the common stock of knowledge*". (Merton 1957, 639).

An underlying assumption here is that these original contributions to science - especially those standing out - are sooner or later recognised by a variety of authoritative actors in science systems and beyond. We argue that those recognitions leave trails, which we can follow to identify those advances, which stand out.

Figure 1 briefly describes this approach.

Millions of scientific papers are published every year and some of them are recorded by bibliographic services such as libraries, registration agencies or bibliometric databases (which usually are selective of sources they index).

Only some of them (few?) stand out by the contribution they make to the existing body of knowledge and are recognised as such. We identified five groups of actors in the science communication system which are engaged in this recognition and whose trails can be followed (relatively) easily.

**Figure 1: Recognition Channels and Trails**

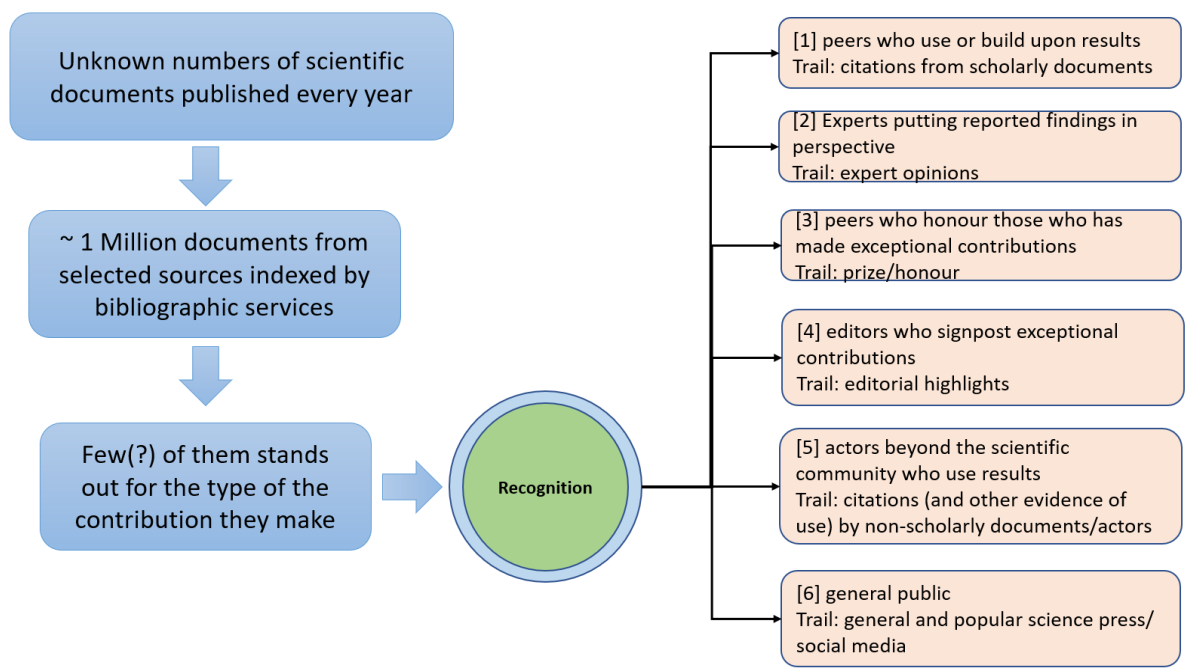

The traditional bibliometric approach uses citation data to this end. It posits that fellow researchers who are inspired by the work or use or build upon it would acknowledge their debt by citing it. The accumulated citations are therefore a form of recognition of some sort. Admittedly there are discussions of what citations really measure, but no one doubts that some of the citations reflect the fact that the results of the cited work have been of value for the citing work. It is in this sense considered here as a form of recognition.

The recognition of peers extends however beyond citing the work. Other actors also have their ways to recognise the work.

- The scientific community and other organisations such as philanthropic foundations also often recognise scientific advances in the form of prizes and other types of accolades that they bestow upon those who are considered to have made exceptional advances.

- Experts who comment (often by invitation) on reported advances (single finding or a collection of related advances), putting it in perspective to assess its potential significance.
- Journal editors who, depending on their format and editorial policies, have also different ways of signposting advances which stand out.
- Non-scientific actors - such as regulatory bodies (or public services designing/enacting policies)- who also use scientific results in their work also usually acknowledge it by citing it.
- The general public whose imagination is often captured by scientific advances. Traditionally the general public is informed by general and popular science press and increasingly rely on social media for scientific information.

It is clear that the different recognition forms (and even within the same categories) will emphasise different aspects in which a paper "stands out". We purposely use the terms "outstanding advances" to make this point clear. For example, researchers may cite a paper because they use its data or methods or because researchers build upon insights it reports.

Citation rates in this case would reflect the relevance of the cited work. Equally, prizes and honours may emphasise a different aspect such as the potential of a discovery to generate tangible societal benefits.

Ultimately it remains a research question for the bibliometric community to understand and document those recognition channels together with their main characteristics. There is a growing literature, for example, on how different forms of recognition correlate with each other. To name just a few of the recent studies:

- Antonoyiannakis, M. (2015 and 2017) found that papers highlighted as "editor's choice" by Physical Review B outperform comparable papers in terms of citation counts.
- Wainer et al (2015) collected citation data of papers which received "best paper awards" at prestigious computer science conferences and found them to be on average more cited than comparable papers.
- Mutz et al (2016) analyzed the citation patterns of the papers designated as "very important paper" (VIP) in Angewandte Chemie International Edition and found that they had a significantly higher citation performance than comparable papers identified through an elaborated propensity score marching approach..
- Bornmann and Tekles (2020) used the list of milestone papers published in journal Physical Review Letters (PRL) (39 papers from over 44,000 articles published between 1980 and 2002) to validate several research disruptiveness indicators.
- Wuestman et al. (2020a) used the list of breakthroughs of the year of Science magazine to study their characteristics. By qualitatively analysing the content of the reference papers, they find that "*most scientific breakthroughs are driven by a question and in line with literature, and that paradigm shifting discoveries are rare.*" (p. 1)

A major obstacle to this type of inquiry, however, is the lack of readily available datasets – with the exception of citation data - recording which papers have been "recognised" by which channels. This can explain also the fact that most of those studies listed above use a single dataset.

A partial objective of this paper is thus to trigger discussions in the bibliometric communities on how those datasets can be created, distributed openly and collectively maintained.

Our focus in this paper is on "editorial highlights", for which we present a prototypical exercise to collect and curate data.

## 3. Editorial highlights

*3.1 The scope*

Editors of scholarly journals play a central role in the scientific communication system. Their main responsibility is to oversee the assessment of the manuscripts submitted by authors and make decisions on their publications. As such they are "gatekeepers of knowledge" in their respective fields. In the course of their work, editors get acquainted with the main trends and development of their fields, not only through the manuscripts they received but also through the comments of the reviewers and –presumably – an active interest in the field and other journals.

Sometimes, editors of scholarly journals put this knowledge to use by identifying and making public what they think has been important discoveries and developments in the areas covered by their journal.

We can distinguish between two types of highlights by journal editors: (a) highlighting papers within their own journals (e.g. among a selection among those published in the previous year) and (b) highlighting publications in other, different journals. We refer to the former as "*inward editorial highlights*" and the later as "*outward editorial highlights*".

*3.1.1 Inward editorial highlights*

The most common form in which journal editors highlight what they see as the most important publications in their journals are the "best of" annual lists. It is not always clear which criteria are used to select entries: in some cases, it is editorial judgments in other cases bibliometric indicators such as number of downloads and sometimes it is a combination of both. An example of the best of collection can be found at Cell Press (https://www.cell.com/bestof).

The recognition of most important papers in own journals can also be in the form of Prizes. For example, PNAS has the Cozarelli Prize, given annually to 6 papers, whose PNAS published work "represent exceptional scientific achievement, originality, and innovation in their fields". The AAAS Newcomb Cleveland Prize recognizes papers published in Science which represent "a fundamental contribution to basic knowledge or is a technical achievement of far-reaching consequence".

Some journals also have a "post-publication peer review process" by which they invite domain experts to comment on a given paper, putting in perspective the advance it reports and the implications. Examples of journals using this form include Science ("Perspective") and Nature ("News and Views"). Those can also be "expert opinions".

*3.1.2 Outward editorial highlights*

This category of editorial highlights refers to editors highlighting important publications from a broader pool of publications beyond the journals' own papers.

It should be noted that the task of identifying and communicating on most important papers is not unique to editors of scholarly journals. There are two additional categories of editors who play an important role in this respect: editors of popular science press and editors of encyclopaedia yearbooks.

Together with books, lectures, exhibitions etc. the popular science press aims to increase the science literacy of the general public by communicating scientific advances and other related news to the general audience.

- The editors of the *popular science press* sift through the mass of published advances and select those they judge to be of interest to their readers. As shown by the prevalence of the "news you can use " type of articles in popular science press, the choice is dedicated mainly by the communication of news with practical consequences/utility to the readers or those which catch the public imagination. Some of those journals however also publish lists of discoveries they consider most important in a given time frame (usually a year).
- *General audience encyclopaedias* aim to present the summary of human knowledge in a systematic and comprehensive manner. Traditionally all major encyclopaedias have had annual supplements (generally in forms of yearbooks) which highlight major events in the previous year, including major advances in sciences.

In our view, all three types of editors are authoritative sources of information on important advances as they witness first-hand scientific advances as they happen and select those, they consider most important to highlight. The selection of important discoveries they publish is what we refer to here as "outwards editorial highlights" - the signaling of important/interesting advances in a given research field by editors of scholarly journals, renown popular science press and encyclopaedia editors. A unifying feature of the "outwards editorial highlights" as defined here is the fact the editors choose the advances to highlight from a broader pool of published results.

**Table 1: Selected Outwards highlights**

| Annual lists : discoveries of the year | | |
|---|---|---|
| 1 | Science - Breakthrough of the Year | given annually by Science to the "most significant development in scientific research" since 1996. According to wikipedia, it has been inspired by the Time's Man of the Year. With every Breakthrough of the Year, Science publishes also the runners up (generally about 9). |
| 2 | La Recherche - Découvertes de l'année | A list of top 10 discoveries of the year. It is usually published in January covering the discoveries / developments published in the preceding year. |
| 3 | Physics World - Breakthrough of the Year | From the discoveries reported about in a given year, Editors of Physics World short list top 10 advances using those criteria:(a) Significant advance in knowledge or understanding, (b) Importance of work for scientific progress and/or development of real-world applications and (c) of general interest to Physics World readers. Out of those 10, one is subsequently selected as "Breakthrough of the Year" while the remaining are listed under runners-up |
| 4 | Discover magazine | In a series called "The Year in Science", Discovery magazine publishes in its Jan/February issue the top stories from the previous year. |
| **Encyclopedia Yearbooks/Lists** | | |
| 1 | Encyclopaedia Universalis - La Science au présent | Since 1997, in addition to the Yearbook, Encyclopaedia Universalis publishes every year a book called "Science au présent". Its structure has changed over time. In recent issues, important scientific advances from the previous year are presented in a section called "Etapes" (qui) "*font le point sur les avancées récentes qui viennent compléter nos connaissances, voire les bousculer*" |
| 2 | Wikipedia – Year in Science Series | Year in science lists significant scientific events that occurred in the year (or are scheduled to occur in case of on-going year). The oldest entry is seem to go back to the year 1500. |
| **Regular lists (on a more frequent basis)** | | |
| 1 | Science – Editors' Choice | Every week, Science publishes a selection of highlights summarizing researchers advances reported in other journals |
| 2 | Nature - Research highlights | A weekly selection by Nature Editors of highlights from the literature from "other journals". (Some other nature journals have also the editorial highlights which follow their periodicity) |

*3.2. Identification and collection of outward editorial highlights*

The focus of this paper is on "outward editorial highlights". In the following sections, we discuss the collection of data for two of three categories: annual lists and the regular lists. The category "Encyclopaedia Yearbooks" will be discussed separately.

*3.2.1 Annual lists / discovery of the year*

Looking at the venues which publish the outward editorial highlights as defined here, we could identify several series which are presented below.

The list is not meant to be exhaustive but rather a starting point for reflections on how the data on the advances can be gathered and made easily accessible for bibliometric analysis. The table makes a distinction between three categories: (1) annual lists which include Science's "Breakthrough of the Year" and three annual series of popular science magazines; (2) encyclopaedia yearbooks/ with two entries: Encyclopaedia Universalis' "Science au Présent" and "Wikipedia Year in Science"; and (3) other lists which are published on more frequent basis. In these categories, we could identify Science's "Editor's choice" and "Research highlights" from Nature Journals.

To the best of our knowledge, there is currently no service, which systematically collects data on which discoveries/papers are subject to editorial highlights.

We prototyped a workflow for data collection and curation of selected editorial highlights. The first step is straightforward since it is just about collecting the list of the discoveries and their URLs. In this case we collected all entries from the original lists including those which refer to major events in science systems.

The objective of the second step is to record an original research paper reporting the discovery. This proved to be slightly more laborious. While most of the stories on the discovery of the year included a reference list, in some cases extra work was needed where: (1) some references were provided simply as string and identifiers (doi) needed to be found; (2) some reference lists included not only original research but also some background material, reviews and news articles; or (3) in the relatively rare instances where there are no reference lists at all (or all references point to news articles). In this latter case, the original articles have to be inferred either from the discovery of the year article or news articles on the discoveries.

*3.2.2 Regular lists*

Regarding the editorial highlights published on a more regular basis (as opposed to annual lists) we focused on two sources: Nature and Science Magazines.

Since 2017, Springer Nature has created SN SciGraph: an open dataset with SN publications linked between them and linked also to a variety of other datasets. To our knowledge however, the research highlights published weekly in Nature Magazine or regularly by other Nature journals are not recorded individually and linked to original research papers. Nature and Science "highlights" therefore had to be assembled from scratch.

The collection from both sources proved tedious as it involved creating the dataset from the information available from their websites (especially finding the identifiers of the papers mentioned in the highlight in Science magazine - "called editor's choice) is difficult as no doi link is offered, only references in classical bibliographic formats).

*3.3 Dataset*

From the four sources listed above, over 8000 primary research papers were identified as being "outward editorial highlights" (i.e. underpinning or closely related to one of several outstanding research advances as highlighted by the editors). Table 2 gives an overview broken down by source.

We note that in the case of the annual lists not all entries correspond to specific research advances reported in the indicated year. Some entries may refer also to important events in science in the year. For example, the 2018 edition of Science Breakthrough also included #MeToo, a movement to end sexual harassment.

Several formats to release the dataset were considered. For this edition, we opted to release it in four spreadsheet files (Microsoft excel) each corresponding to a given source whereby the individual sheets bundle data for a given year. This helps avoiding unnecessary complexity in data management (storage and updating) while facilitating usage. The discoveries of the year (Science's Breakthrough of the year and *les 10 découvertes de l'année*) include also an extra sheet listing the individual entries as provided by source.

Annex 1 provides a brief description of the columns used for each dataset.

**Table 2: Overview of collected data**

| Editorial Highlight | Period | number of entries | number of references resolved |
|---|---|---|---|
| Science - Breakthroughs of the Year | 2009-2019 | 110 | 292 |
| La Recherche - les 10 découvertes de l'année | 2008-2019 | 120 | 349 |
| Nature - Research Highlights | 2008-2020 | 4,952 | 4,813 |
| Science - Editors' choice | 2009-2020 | 3,800 | 3,243 |

## 4. Summary and Outlook

The bibliometric research community has devoted significant effort to develop approaches to identify a set of publications which represent major advances in their fields, making high impact beyond their areas. The term "breakthrough" is often used in this context. We argued that since the term is rooted in a model of scientific growth, which has been conceptually and empirically challenged, it may mislead more than it can fruitfully guide the effort to detect important advances.

From the obvious observation that research advances are "important" because some - authoritative - actors in the research communication system recognized them as such (from very different perspectives) we bring to discussion the idea of creating open datasets of different "recognition channels". Focusing on "editorial highlights", the signaling of important/interesting advances (selected from a broader pool of published results) by editors of scholarly journals, renowned popular science press and encyclopaedia editors, we prototyped the creation of an open dataset. The dataset covers the last ten years and includes (a) the "discoveries of the year" namely: Science - Breakthroughs of the year and La Recherche - les 10 découvertes de l'année and (b) weekly editorial highlights from Nature ("research highlights") and Science ("editor's choice").

The final dataset includes 230 entries in the "discoveries of the years" (with over 720 references) and about 9,000 weekly highlights (with over 8,000 references). The result of the gathering data on editorial highlights are encouraging. Although it remains largely a manual task, some of the steps can be automated.

We see three major challenges lying ahead.

- how to expand the dataset by including more sources.
- how to sustain this effort and maintain an up-to-date list of discoveries and related publications.
- how to model the data such that they can be integrated into existing Research Information Graphs.

We note that Wikipedia has an entry on the "Breakthrough of the Year" of Science Magazine listing all the yearly breakthroughs and runners-up since 1996. A closer look at the entries shows, however, minor problems of accuracy (some entries in Wikipedia lists could not be matched to Science lists) and missing references (Wikipedia has no references to the original papers reporting the discoveries).

A possible enhancement would be to fix those minor problems and to add other annual lists which in our view are of encyclopedic relevance such as "les découvertes de l'année" by La Recherche and Physics World Breakthrough of the year. This would make Wikipedia the authoritative sources of the discoveries of the year. The implementation of this idea is of course subject to a positive reaction of Wikipedia editors and the Wikipedia community. A suitable format would also have to be agreed.

The next step (and challenge) is set up a decentralized and collaborative effort to source new venues of editorial highlights, keep the lists updated and develop data models suitable for Open Research Information Graphs, organized as a decentralized collective and collaborative effort.

Such graphs could also include similar or related datasets produced in course which are made available by researchers such as the list of publications by Nobel Prize laureates (Li et al 2019) or the characteristics of discoveries included among Science magazine annual list of Breakthroughs of the year (Wuestman et al. 2020b).

**Acknowledgment**

We would like to thank following people who contributed in a variety of ways on the work presented. Anthi Kyriakou, Andreas Lang both trainees at the ERCEA in 2014 and 2016 respectively helped in conceptual design, exploratory data collection and design of data flow. Discussions with Manolis Antonoyiannakis, Scientific Advisor to the ERC President (2008-2010) helped in the conceptual phase. Lutz Bornmann and Ludo Waltman provided helpful insights.